\documentclass[12pt]{article}
\usepackage{graphicx}

\begin{document}

 \title{Intrinsic dissipation in cantilevers}
\author{G.P. Berman $^{1}$\footnote{Corresponding
author: gpb@lanl.gov} and A.A. Chumak$^{1,2}$
\\[3mm]$^1$
 Los Alamos National Laboratory, Theoretical Division,\\ Los Alamos,
NM 87545
\\[5mm] $^2$  Institute of Physics of
the National Academy of Sciences\\ pr. Nauki 46, Kiev-28, MSP
03028 Ukraine
\\[3mm]\rule{0cm}{1pt}}
\maketitle \markright{right_head}

\begin{abstract}

We consider the effects of a velocity-independent friction force
on cantilever damping. It is shown that this dissipation mechanism
causes nonlinear effects in the cantilever vibrations. The size of the nonlinearity
increases with decreasing cantilever velocity. Our analysis makes 
it possible to understand Stipe's $\it et$ $\it al.$ \cite{ru} experiments where an 
amplitude dependence of the cantilever eigenfrequency and anomalous dissipation was 
observed only at small amplitudes.

\end{abstract}

\section{Introduction}
The quality and potential practical opportunities for
micromechanical devices strongly depend on
their dissipative characteristics. Numerous authors have studied the
damping of cantilever vibrations both theoretically and
experimentally. There are at least two obvious motivations for these
studies. First, the amount of dissipation determines the sensitivity of
cantilever-based devices. Hence, reducing dissipation
improves the technical characteristics of the devices. The second is
based on the possibility of probing the dissipative forces which arise
due to cantilever-sample interaction in order to obtain a useful
information about the sample. The last idea was a basic one for
originating a new direction for atomic force microscopy: 
noncontact dissipation force microscopy. (See references
\cite{gots}-\cite{ricar}).

Many attempts to understand the dissipation mechanisms in
cantilevers were undertaken. The corresponding theoretical models,
which could describe cantilever damping,  are presented in
references \cite{roukes} and \cite{roukes1}. At the same time, the
fundamental nature of damping is still not clear, especially in the
case of noncontact cantilever vibrations. (See, for example,
discussions in references \cite{ru},\cite{gots},\cite{ricar}, and
\cite{chu}). Fluctuations of the van der Waals force (``vacuum
friction") or Joule losses are not responsible for the dissipation
measured experimentally. The corresponding friction effects have
been calculated in \cite{vol}-\cite{chu} to be many orders of
magnitude smaller than those measured in \cite{ru}. There is not
only quantitative disagreement, but also different qualitative
behavior. The point is that both theories predict very strong
dependence of damping on the cantilever-sample separation, and this
is not consistent with the experimental observations.

A more complex and more specific model, involving adsorbed
particles, was suggested in \cite{vol05}. The authors connect the
large long-range noncontact friction with the electromagnetic
interaction of moving charges, induced on the surface of the tip by
the bias voltage, with acoustic vibrations in an adsorbate layer on
the surface of the sample.

An internal quantum friction due to the presence of a finite number
of two-level systems is analyzed in  Refs. \cite{gaid} and
\cite{neto}. Their analysis is related to millikelvin temperatures
which is beyond experimental conditions of \cite{ru}.

The experimental dependence of the friction force, $F$, on the bias
voltage, $V$, applied between the tip and a sample, has been found.
It is the quadratic dependence, $F\sim V^2$ (see, for example,
references \cite{ru} and \cite{ricar}) that is really typical of the
Joule mechanism for energy losses. At the same time, it should be
emphasized that not only Joule losses, but also a tip-sample
attractive forces, caused by electrostatic interactions between
closely spaced charged surfaces, behave as $V^2$. Taking into
account the fact that the Joule mechanism is not effective (at least for good
conductors), the increased dissipation may be attributed to some
effects of the attractive forces. In this connection, it should be
noted that the increase of dissipation with decreasing 
tip-sample spacing was also obtained in the absence of the
electrostatic attraction (for zero-value bias-voltage). The
calculations of \cite{chu} show that at small separations the van
der Waals (Casimir) attractive force may become so strong as to be
able to modify the cantilever eigenfrequencies.

A very specific attraction of the metal tip to dielectric samples occurs as well.
Although overall the dielectric sample is electrically neutral, it
may contain localized electric charges, which
interact with the corresponding ``image" charges in the tip. Samples of fused
silica were used by authors of \cite{ru} to study the
effect of charged centers on the cantilever vibrations. (Reference \cite{chu} contains
the corresponding theoretical calculations.)  The number of centers has been
varied by means of irradiation with $\gamma$ rays. An
enhancement of dissipation occurs when the concentration of charged defects increases.

The increasing attractive forces in the above systems are accompanied by
the growth of dissipation. Although this force may be of a different physical nature,
it always results in an increase in cantilever dissipation. Therefore any consistent
theory of dissipation should take into account this circumstance.

Concluding the introduction, we highlight a statistical
analysis of various micromechanical systems \cite{roukes1} that shows
that the quality factor (inverse of dissipation) usually scales
linearly with size. There is a general tendency of dissipation to
increase with the surface/volume ratio, indicating the relevance  of
the cantilever surface layers to dissipation.

\section{The friction force}

In general, the cantilever motion is described by the oscillator model. The
corresponding equation of motion is given by
\begin{equation}\label{one}
 m\frac {\partial ^2X}{\partial t^2}+
 \Gamma \frac {\partial X}{\partial t}+kX=0,
\end{equation}
where $X=X(t)$ is the displacement of the cantilever tip; $m$ and
$k$ are the cantilever effective mass and spring constants,
respectively. The term $\Gamma \frac {\partial X}{\partial t}$ is
the friction force, which is usually assumed to be proportional to
the velocity of the cantilever motion. Hence, this force tends to
zero when $\frac {\partial X}{\partial t}\rightarrow 0$. The
experiments in \cite{ru} were performed for the cantilever frequency
$\omega_0=(k/m)^{1/2}\approx 24\times 10^3s^{-1}$ and vibration
amplitudes of the order of $10nm$ or less. This corresponds to the
characteristic value of the velocity equal to $0.02cm/s$. This slow
cantilever motion, in practice, excludes dissipation mechanisms due
to both Joule losses and fluctuations of the van der Waals forces.
Therefore, it seems to be quite reasonable to consider a
velocity-independent friction force, $F$, as the one responsible for
dissipation in this case. The conventional term
``velocity-independent force" means a force which does not depend on
the absolute value of the velocity but always acts in the
direction opposite to the tip velocity. Thus, $F$ reverses its sign
after each half-period of vibrations. This force occurs in
mechanical systems in which one solid surface is in contact with another
and slides along it.

The friction force is proportional to the effective contact area,
which, depends linearly on the compressing force (the
Coulomb friction law). A very exotic mechanism of velocity-independent
friction due to photon exchange between surfaces which are in
close proximity to one another (but not in a contact) was predicted
in Ref. \cite{pen}.

For the case of a cantilever \cite{ru} whose body has a complex
multilayer structure, an internal friction force between the different
layers may arise. The body is a platinum-coated single-crystal
silicon beam. There is also a $1 nm$ titanium layer between the $10 nm$
platinum ``coat" and the silicon. Besides that, the tip was coated with
$200nm$ of gold by evaporation. It was established experimentally that
the dissipation of the platinum-coated single-crystal silicon cantilever
is higher by a factor of $6$  than that of an uncoated cantilever.
Therefore, a muli-layer structure is essential in forming the
overall dissipation. It seems to be reasonable to consider a
tip-sample stretching force to be responsible for increasing the contact
interlayer area and, consequently, the frictional force, both of
which linearly depend  on the tip-sample attraction.

The cantilever motion is described by the equation
\begin{equation}\label{two}
m\frac {\partial ^2X}{\partial t^2}+kX=F,
\end{equation}
where $F>0$ for $\frac {\partial X}{\partial t}<0$ and $F<0$ for
$\frac {\partial X}{\partial t}>0$. Formally, Eq.~\ref{two} may be
rewritten in the form given by Eq.~\ref{one}, with the coefficient $\Gamma$
depending on the velocity:
\begin{equation}\label{three}
\Gamma =|F|{\bigg |}\frac {\partial X}{\partial t}{\bigg |}^{-1}.
\end{equation}
The oscillator equation with $\Gamma$ given by Eq. \ref{three} is
no longer a linear equation. Nevertheless, it can be easily solved for
each time interval between any two successive turning points.
The general solution is given by
\begin{equation}\label{four}
X(t)=(-1)^{n+1}\frac {|F|}{k}+[-X_0+\frac {|F|}{k}(2n+1)]cos(\omega_0t),
\end{equation}
where $n$ is the integer part of $\omega_0t/\pi$ and $-X_0<0$ is
the initial position of the tip.
\begin{figure}[t]
\centering
\includegraphics{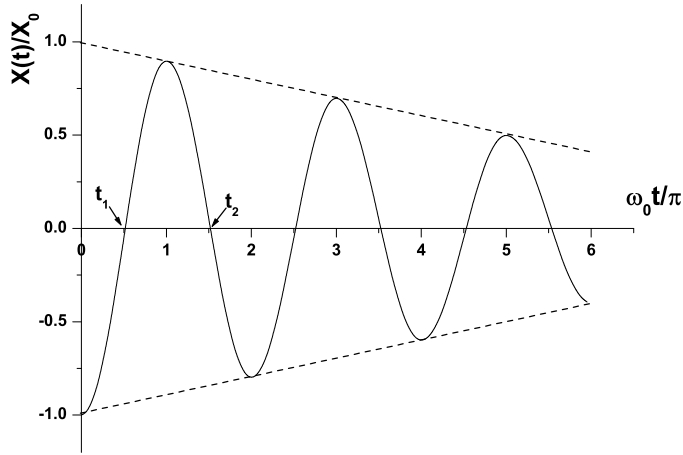}
\caption{Ringdown oscillations in the case of a velocity-independent force.
Points, where  $X=0$, are slightly displaced from the values $n+\frac 12$ which
would correspond to harmonic oscillations.}
 \end{figure}
Eq.~\ref{four} describes the decaying motion of the cantilever. In the
range of small dissipation (the energy loss per period is small 
compared with the total energy $\varepsilon $) and small values of
time, Eq.~\ref{four} may be approximated by a simplified formula
 \begin{equation}\label{five}
X(t)\approx -X_0(1-\alpha t)cos(\omega_0t)\approx -X_0e^{-\alpha t
}cos(\omega_0t),
\end{equation}
where
 \begin{equation}\label{five1}
\alpha\equiv \alpha (X_0)=\frac {2|F|}{\pi m\omega_0X_0}.
\end{equation}
The coefficient $\alpha$ describes the nonlinear damping of the
oscillations, and it depends strongly
on the amplitude $X_0$. As we see, the nonlinearity is the
most pronounced for small amplitudes as was observed
experimentally \cite{ru} (at amplitudes of the order of or less
than $10nm$). It was noticed in Ref. \cite{ru} that this effect
vanishes if either the amplitude of oscillation or the tip-sample separation
increases.

 With the known oscillator trajectory described by Eq. (\ref{four}), 
it is easy to estimate the value of
nonlinear shift of the oscillation frequency. In what follows we
consider the so called ``ringdown" regime of oscillations, which
takes place when a drive circuit is abruptly grounded and the
cantilever rings down until thermal equilibrium is established. The
value of the cantilever frequency may be obtained by measuring the
time interval $t_2-t_1$ between two successive crossing points where
the displacement $X$ is equal to zero. For our model, these points
do not match with the values $\pi/2\omega _0$ and $3\pi/2\omega _0$
as would occur for harmonic oscillations (see Fig. 1), but are
slightly displaced to greater values of $t$.
Considering these displacements as small quantities, we can easily
obtain the difference $t_2-t_1$.

Let us introduce the notation, $\omega_0 t_1\equiv \frac \pi2 +\delta _1$ and
$\omega_0 t_2\equiv \frac {3\pi}2 +\delta _2$. Then we obtain from Eq. (\ref{four})
two equations for $\delta _1$ and $\delta _2$. They are 
\[\frac {|F|}k+\bigg (-X_0+ \frac {|F|}k\bigg )sin\delta _1=0 and\]
\begin{equation}\label{a1}
\frac {|F|}k+\bigg (-X_0+ 3\frac {|F|}k\bigg )sin\delta _2=0.
\end{equation}
It follows from Eqs. (\ref{a1}) that  $\delta_1 \approx \frac {|F|}{kX_0}
\bigg (1+\frac {|F|}{kX_0}\bigg )$ and $\delta_2 \approx \frac {|F|}{kX_0}
\bigg (1+3\frac {|F|}{kX_0}\bigg )$.
Then we have
 \begin{equation}\label{six}
t_2-t_1={\pi \over \omega_0}+{2F^2\over \omega_0k^2X_0^2}.
\end{equation}
The corresponding value of the frequency is given by
\begin{equation}\label{seven}
\omega\approx {\pi \over t_2-t_1}\approx \omega_0{\bigg(1-{2F^2\over
\pi k^2X_0^2}\bigg)}=\omega_0 \bigg(1-{F^2\over \pi
k\varepsilon}\bigg),
\end{equation}
where the energy of vibrations, $\varepsilon $, is equal to $kX_0^2/2$.

The experimentally observed decrease of the frequency is illustrated in Fig. 2d of
reference \cite{ru}. The decrease in the oscillation frequency takes place at the
beginning of the transition from high- to small-amplitude regimes of
oscillations. It would be interesting to study experimentally the dependence of the
frequency shift on the tip-sample attraction, $F_{attr}$. For the case of the
Coulomb law for the friction force, i.e. when $|F|=c_0+c_1F_{attr}$ ($c_{0,1}$
are costants), the frequency shift is given by the expression
\begin{equation}\label{a2}
\Delta \omega\approx -\frac {\omega _0}{\pi k\varepsilon}\bigg(c_0+c_1F_{attr}
\bigg )^2.
\end{equation}
When the attraction force is induced by the bias voltage, $V$, $F_{attr}\sim V^2$.
When it is due to localized charges in the dielectric sample, $F_{attr}\sim n_{ch}$,
where $n_{ch}$ is the concentration of charges. In both cases the model
proposed here can be examined experimentally.

There is an alternative method of calculation of the frequency shift. It is based
on the asymptotic theory of nonlinear oscillations developed by Bogolubov and
Mitropolskii in \cite{bo}. The direct application of this method to our specific case of
a velocity-independent friction force
results in a frequency shift which is slightly smaller (by factor of $3/\pi $) than that
given by Eq. \ref{seven}. The details are in Appendix A.

So far our analysis assumed a constant
friction force. This force may be considered as originating from many
discrete events of energy losses. For brevity, we will use the
term ``kick" for each such event. The duration of each kick is assumed
to be very short compared with the period of vibrations, and the
number of kicks per period is large. After averaging over a time interval much
shorter than the period of oscillations, but  large enough to include many kicks,
the effect
of kicks reduces to a constant friction force. It is evident
that the description based on a continuous friction force is applicable to
high-amplitude oscillations only. The discrete nature of the dissipation
may reveal itself at small amplitudes when the mean ``free path" $l$ is
longer than or of the same order as the amplitude. The correspondence
of the two models to each other imposes a relation between the parameters of both
$|F|=\Delta \epsilon/l$,
where $\Delta \epsilon$ is the energy loss in the course of each
kick ($\Delta \epsilon <<\varepsilon$).

In what follows, we consider the case of a low-probability kick per
oscillation period. This is possible when $X_0<<l$. If the kick
occurs during the motion away from (towards) the center, the time
of returning to the central position ($X=0$) is decreased (increased). Hence,
the period of oscillations is a fluctuating
quantity. It can be shown that these fluctuations result in a
positive frequency shift that is in contrast to the case of large
amplitudes, $X_0>>l$. The following simple analysis makes it possible to
obtain the value of the shift in an explicit form. As before, we proceed from the
equation of motion for a harmonic oscillator
\begin{equation}\label{eight}
X(t)=X_0cos(\omega _0t),
\end{equation}
which describes the vibrations in the absence of dissipation.
Initially, the oscillator is in the position $X(t)=X_0$. Let us
assume that the kick occurs at
time $t_1$. Then $X(t_1)\equiv X_1= X_0cos(\omega _0t_1)$. The value of $t_1$
can be expressed via $X_1$ as
\[t_1=\frac 1{\omega _0}cos^{-1}\bigg (\frac {X_1}{X_0}\bigg ).\]
After this event, the oscillator continues its motion with a
modified amplitude $C$, which can be obtained from energy conservation. It is
given by
\begin{equation}\label{nine}
C=\sqrt{X_0^2-2\Delta\varepsilon /k}.
\end{equation}
For simplicity, we will consider that $\Delta \epsilon<<\epsilon$.

The modified equation of motion is given by
\begin{equation}\label{ten}
X(t)=Ccos(\omega _0t+\varphi ),
\end{equation}
where $\varphi$ is the phase variation introduced by the kick. It is convenient to introduce
the new time variable, $\tau =t+\varphi /\omega _0$. It can be easily seen that $\tau$ varies
from $\frac 1{\omega _0}cos^{-1}\frac {X_1}C$ (the instant of kick) to
$\frac \pi{2\omega _0}$ (the instant of crossing the $x$ axes). Then the overall time
required to get to the central position $X=0$ is given by
\[ t_1+{\pi\over 2\omega_0}-\frac 1{\omega _0}cos^{-1}\frac {X_1}C\equiv
{\pi\over 2\omega_0}+\tau^+,\] where
\begin{equation}\label{ele}
\tau^+={1\over \omega_0}\bigg (cos^{-1} {X_1\over X_0}-cos^{-1} {X_1\over C}\bigg ).
\end{equation}

It follows from Eq. \ref{ele} that the time required to reach the
center has increased due to the kick. In contrast, the transit time decreases
when the oscillator moves away from
the center. The corresponding time is given by ${\pi \over
2\omega_0}+\tau^-$, where
\begin{equation}\label{twe}
\tau^-={1\over \omega_0}\bigg (sin^{-1} {X_1\over X_0}-sin^{-1}
{X_1\over C}\bigg ).
\end{equation}
As we see the variation of the transit time depends essentially on the
position, $X_1$, of the oscillator at the instant of kick. We
assume that all realizations of $X_1$ have equal probabilities.
This assumption will make it possible to take into account the
contribution of the transit-time fluctuations to the frequency shift.

As before, our following analysis concerns the case of the ringdown
experiment. The time required to undertake $N$ full cycles can be expressd as
\begin{equation}\label{thi}
t=N{2\pi\over \omega_0}(1+\Delta),
\end{equation}
where
\[\Delta =\frac 1{2\pi N}\bigg [\sum _i\tau ^-(X_i)n_i+\sum _j\tau ^+(X_j)n_j \bigg ].\]
The indices $i$ and $j$ denote the quarter-periods with away-from-center and
towards-center motion, respectively. The random
variables $n_{i,j}$ are equal to $1$ or $0$ depending on whether or not a kick occurs during
the ($i,j$)th  quarter-period. The average value of $n_i$ is given by
$\langle n_i\rangle=\frac {X_0}l<<1$, implying a negligible probability of two kicks
during any given quarter-period. The period of oscillations,
$T$, is defined as $T=t/N$. Hence, the frequency of oscillations is
given by
\begin{equation}\label{fort}
\omega =2\pi N /t=\omega_0(1+\Delta)^{-1}\approx \omega _0(1-\Delta +\Delta ^2-...).
\end{equation}
After statistical averaging of $\omega$, the linear term $\langle \Delta\rangle$ vanishes.
This follows from the definitions of $\tau^{\pm }$ and the trigonometry relation
$sin^{-1}(x)=\frac \pi 2-cos^{-1}(x)$. The main contribution to the frequency shift
is from the term $\langle \Delta ^2\rangle $. Thus, we have
\begin{equation}\label{fift}
\langle\omega \rangle -\omega _0\approx \frac {\omega _0\langle n\rangle }{\pi ^2N}
\langle \big [\tau ^-(X_i)\big ]^2\rangle .
\end{equation}
In deriving Eq. \ref{sixt}, we use the following relations: (i)
$n_i^2\equiv n_i$, (ii) $\langle n_in_j\rangle=\langle
n_i\rangle\langle n_j\rangle$ when $i\not=j$. Also, the integer $N$
cannot be very large because the energy dissipation during the
observation time is assumed to be small.

Considering the losses as statistically independent events which have equal probabilities
at any value of $X$ within the interval $[-C,+C]$ (the interval where a kick
is possible), we can calculate the average value of $[\tau ^-]^2$ as
\begin{equation}\label{fift1}
\langle [\tau ^-]^2\rangle =\frac 1C\int _0^CdX\bigg [sin^{-1}\bigg(\frac XC\bigg )-
sin^{-1}\bigg (\frac X{X_0}\bigg )\bigg ]^2.
\end{equation}

This integral can be calculated analytically considering $\frac {\Delta \varepsilon}
{2\varepsilon }<<1$. (See Appendix B). Finally, we get
\begin{equation}\label{sixt}
\langle\omega \rangle -\omega_0\approx {|F|\Delta \varepsilon \over
4\sqrt{2}\pi^2N\sqrt{m}\varepsilon ^{3/2}}ln\bigg ({2\varepsilon \over
\Delta \varepsilon}\bigg ).
\end{equation}

Comparing the amplitude dependence of the frequency shift in two
limiting cases given by Eqs. \ref{seven} and \ref{sixt}, we may
conclude that the function $\omega (X_0)$ has a minimum at intermediate
amplitudes as shown in Fig. 2.
\begin{figure}[ht]
\centering
\includegraphics{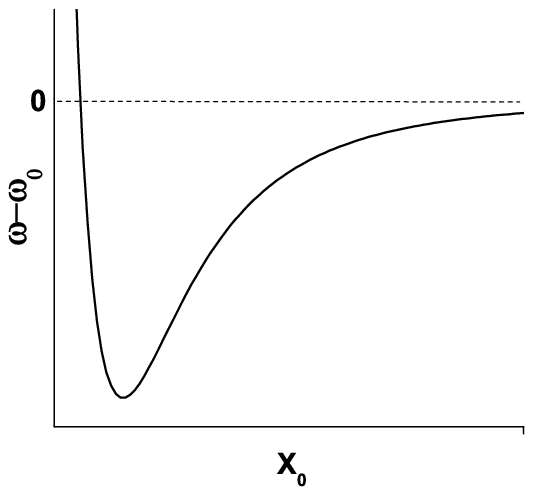}
\caption{Dependence of frequency shift on the amplitude of oscillations $X_0$.}
\end{figure}
 This curve is similar to the experimental one displayed
 in Fig.~2d of reference \cite{ru}.

\section{Conclusion}

We have described phenomenologically a possible mechanism that produces a small-amplitude
nonlinearity for cantilever vibrations. It is assumed that the nonlinearity is due to
velocity-independent friction force which, at small
amplitudes,  reduces to a sequence of random kicks retarding cantilever motion.
Our analysis predicts both the increase
of the decrement of oscillations and  nonmonotonic behavior of the frequency shift when the
amplitude decreases. It is shown that the last effect is due to the crossover from the amplitude-
independent friction force to discrete events of the cantilever energy losses.
This is in a qualitative agreement with the cantilever vibrations observed experimentally.

\section{Acknowledgment}
We thank G.D. Doolen for discussions. This work was carried out
under the auspices of the National Nuclear Security Administration
of the U.S. Department of Energy at Los Alamos National Laboratory
under Contract No. DE-AC52-06NA25396 and the Ukrainian project VTS/138 (Nanophysics).

\begin{appendix}

\section*{Appendix A}

A regular method for solution of the equations
\begin{equation}\label{A1}
{\ddot X}(t)+\omega _0^2X=\epsilon f(X,{\dot X}),
\end{equation}
where $\epsilon $ is a dimensionless quantity ($\epsilon \rightarrow 0$), is developed
in \cite{bo}.  In our case, the term on the right-hand side is given by
\begin{equation}\label{AA1}
\epsilon f(X,{\dot X})\equiv -\frac Fm\frac {\dot X}{|\dot X|}.
\end{equation}
According to \cite{bo}, the solution of Eq. \ref{A1} can be represented as an expansion in powers of
$\epsilon$,
\[{ X}(t)=acos\varphi +\epsilon u_1(a,\varphi )+\epsilon ^2u_2(a,\varphi )+O(\epsilon ^3),\]
\[\dot a=\epsilon A_1(a)+\epsilon ^2A_2(a)+O(\epsilon ^3),\]
\begin{equation}\label{A2}
\dot \varphi =\omega_0+\epsilon B_1(a)+\epsilon ^2B_2(a)+O(\epsilon ^3).
\end{equation}
To obtain the decrerment and frequency shift of oscillations, it is sufficient to calculate
only terms up to the second order in $\epsilon^2$. For the case of the dissipation force
given by Eq. \ref{AA1}, these quantities are determined by
\[A_1(a)=\frac {F_1}{2\omega _0},\]
\[B_1(a)=A_2(a)=0,\]
\begin{equation}\label{A3}
B_2(a)=-\frac {F_1^2}{4\omega_0^3a^2}-\frac 1{2\omega _0^3a^2}\sum_{n=2}^\infty
\frac {n^2F_n^2}{n^2-1},
\end{equation}
where
\[F_n=-\frac {4|f|}{\pi n}sin\bigg(\frac {\pi n}2\bigg)\]
The quantity $A_1$ determines the decay of oscillations. Setting the initial condition for
the amplitude $a$ corresponding to the solution given by Eq. \ref{four}, $a(t=0)=-X_0$, we
obtain from Eqs. \ref{A2} and \ref{A3}
\begin{equation}\label{A4}
a(t)=-X_0\bigg (1-\frac {2Ft}{\pi m\omega _0X_0}\bigg ).
\end{equation}
This coincides with the results given by Eqs. \ref{five} and \ref{five1}.

The value of the renormalized frequency
$\omega (a) =\dot \varphi _(a)$
can be obtained explicitly after summing up in the expression for $B_2(a)$.
This sum is equal to $F_1^2/4$.
Then the frequency of oscillations is given by
\begin{equation}\label{A5}
\omega =\omega_0\bigg (1-\frac {6F^2}{\pi^2k^2X_0^2}\bigg ).
\end{equation}
The second term in the brackets is very close to that in Eq. \ref{seven}. Their ratio is
equal to $3/\pi$ which is close to unity. Thus both approaches give almost identical results.

\end{appendix}
\begin{appendix}
\section*{Appendix B}
\[ \langle [\tau ^-]^2\rangle =\frac 1C\int _0^CdX\bigg [sin^{-1}\bigg(\frac XC\bigg )-
sin^{-1}\bigg (\frac X{X_0}\bigg )\bigg ]^2=\]
\begin{equation}\label{ap1}
\int _0^1dx\big [sin^{-1}x-sin^{-1}x(1-\delta )\big ]^2,
\end{equation}
where $\delta \equiv \Delta \varepsilon /(2\varepsilon )<<1$. Let us set
\[sin^{-1}x(1-\delta )=y+\delta ^\prime, \]
where $|\delta ^\prime | <<1.$ Then we have $siny=x$ and
\[\delta^\prime \approx \sqrt {x^{-2}-1}-\sqrt {x^{-2}-1+2\delta}.\]
After substituting $x=(1+z)^{-1/2}$, the last integral in (\ref{ap1}) is reduced to
\begin{equation}\label{ap2}
\delta^2\int _0^\infty \frac {dz}{(1+z)^{3/2}}\frac 1{z+\delta +
\sqrt {z^2+2z\delta}}.
\end{equation}
One cannot set $\delta =0$ in the integrand of Eq. (\ref{ap2}) because of logarithmic
divergence of the integral.

Let us denote the last integral as $K$. It is convenient to divide
the range of integration in two parts and express the whole integral
as a sum $K=K_1+K_2$, where \[K_1=\int _0^{\sqrt
\delta};~K_2=\int _{\sqrt \delta}^\infty.\] The integrand in $K_1$
can be approximated by $[z+\delta +\sqrt {z^2+2z\delta }]^{-1}$, and
it can be easily seen that $K_1\approx (1/4)|ln\delta|$.

The integrand in $K_2$ can be approximated by $\bigg [(1+z)^{3/2}2z \bigg ]^{-1}$. For
small values of $\delta$, $K_2\approx (1/4)|ln\delta|$. Then, $K\approx (1/2)|ln\delta|$.
Finally, using Eqs. (\ref{ap1}) and (\ref{ap2}), we have
\begin{equation}\label{a3}
\langle [\tau ^-]^2\rangle =\frac {\Delta \varepsilon ^2}{8\varepsilon ^2}
ln \frac {2\varepsilon}{\Delta \varepsilon}.
\end{equation}
\end{appendix}

\newpage \parindent 0 cm \parskip=5mm


\begin{thebibliography}{99}

\bibitem{ru}  B.C.~Stipe, H.J.~Mamin, T.D.~Stowe, T.W.~Kenny, and D.~Rugar,
\newblock Phys. Rev. Lett. {\bf 87}, 096801 (2001).

\bibitem{gots}  B.~Gotsmann and H.~Fuchs
\newblock Phys. Rev. Lett. {\bf 86}, 2597 (2001).

\bibitem{loppa}  C.~Loppacher, R.~Bennewitz, O.~Pfeiffer, M.~Guggisberg,
M.~Bammerlin, S.~Schar,V.~Barwich, A.~Baratoff, and E.~Meyer,
\newblock Phys. Rev. B {\bf 62}, 13674 (2000).

\bibitem{dorprl}  I.~Dorofeyev, H.~Fuchs, G.~Wenning, and B.~Gotsmann,
\newblock Phys. Rev. Lett. {\bf 83}, 2402 (1999).

\bibitem{tsuka}  M.~Gauthier and M.~Tsukada
\newblock Phys. Rev. B {\bf 60}, 11716 (1999).

\bibitem{rugar1} T. D. Stower, T. W. Kenny, D. J. Thomson, and D. Rugar,
\newblock Appl. Phys. Lett. {\bf 75}, 2785 (1999).

\bibitem{ricar}  R.~Garcia and R.~Perez, \newblock Surf. Sci. Rep.
 {\bf 47}, 197 (2002).

\bibitem{roukes}  A.N.~Cleland and M.L.~Roukes,
\newblock J. Appl. Phys. {\bf 92}, 2758 (2002).

\bibitem{roukes1}  P.~Mohanty, D.A.~Harrington,
K.L.~Ekinci, Y.T.~Yang,M.J.~Murphy, and M.L.~Roukes,
\newblock Phys. Rev. B {\bf 66}, 085416 (2002).

\bibitem{vol} A.I.~Volokitin and B.N.J~Persson,
\newblock Phys.Rev. B {\bf 65}, 115419 (2002).

\bibitem{per} B.N.J.~Persson and Z.~Zhang,
\newblock Phys. Rev. B {\bf 57}, 7327 (1998).

\bibitem{volper} A.I.~Volokitin and B.N.J.~Persson,
\newblock J. Phys.: Condens. Matter {\bf 11}, 345 (1999).

\bibitem{chu} A.A.~Chumak, P.W.~Milonni, and G.P.~Berman,
\newblock Phys. Rev. B {\bf 70}, 085407 (2004).

\bibitem{vol05} A.I. Volokitin and B.N.J. Persson,
\newblock Phys. Rev. Lett. {\bf 94}, 086104 (2005);
A.I. Volokitin, B.N.J. Persson, and H. Ueba,
\newblock Phys. Rev. B {\bf 73}, 165423 (2006).

\bibitem{gaid} G. Zolfagharkhani, A. Gaidarzhy, S.B. Shim,
R.L. Badzey, and P. Mohanty,
\newblock Phys. Rev. B {\bf 72}, 224101 (2005).

\bibitem{neto} S. Seoanez, F. Guinela, and A.H.C. Neto,
\newblock arXiv: cond-mat/0611153 v3 (2007).

\bibitem{pen} J.B. Pendry, \newblock J. Phys.: Condens. Matter {\bf 9}, 10301 (1997).

\bibitem{bo} N.N. Bogolubov and Yu.A. Mitropolskii, {\it Asymptotic Methods
in the Theory of Nonlinear Oscillations} (Fiziko-Matemat.
Literature, Moscow, 1958).



\end{thebibliography}
\end{document}